\documentstyle[aps,pre]{revtex}
\begin{document}
\twocolumn[\hsize\textwidth\columnwidth\hsize\csname @twocolumnfalse\endcsname
\draft
\title{STOCHASTIC LATTICE MODELS WITH SEVERAL ABSORBING STATES}
\author{Haye Hinrichsen}
\address{
	Department of Physics of Complex Systems,
        Weizmann Institute, Rehovot 76100, Israel}
\date{October 1996, to appear in {\it Phys. Rev.} {\bf E}}
\maketitle
\begin{abstract}
We study two models with $n$ equivalent absorbing states that generalize
the Domany-Kinzel cellular automaton and the contact process. 
Numerical investigations show that for $n=2$ both models belong to the
same universality class as branching annihilating walks with an even
number of offspring. Unlike previously known models, these models
have no explicit parity conservation~law.
\end{abstract}
\pacs{PACS numbers: 05.70.Jk, 05.70.Ln, 64.60.Ak}] 
\renewcommand{\theparagraph}{\Alph{paragraph}}
\input epsf
%
%
%
%
\def\figdomanykinzel
{\begin{figure}[h]
\epsfxsize=75mm
\epsffile{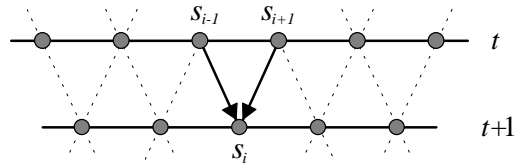}
\caption{Update in the Domany Kinzel model.}
\label{DomanyKinzelFigure}
\end{figure}}
%
%
%
%
\def\figcontactprocess
{\begin{figure}[h]
\begin{center}~
\epsfxsize=75mm
\epsffile{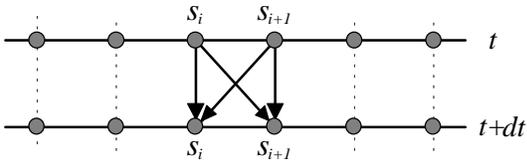}
\end{center}
\vspace{-2mm}
\caption{Update in the contact process.}
\label{ContactProcessFigure}
\end{figure}}
%
%
%
%
%
\def\figphasediag
{\begin{figure}[h]
\epsfxsize=95mm
\epsffile{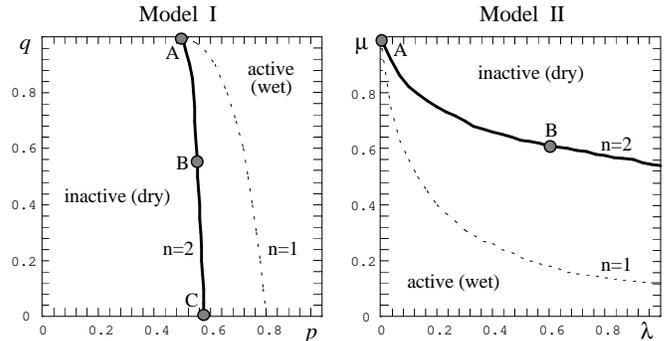}
\caption{The phase diagrams of models I and II for two
absorbing states. The dashed lines indicate the corresponding
transition lines for directed percolation. The explanation of the
points A,B, and C can be found in the text.}
\label{PhaseDiagram}
\end{figure}}
%
%
%
%
%
\def\figsimulations{
\begin{figure}[h]
\epsfxsize=90mm
\epsffile{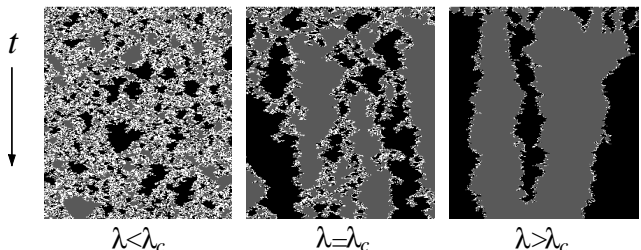}
\caption{Simulation of model II for $n=2$
starting from a random initial condition.
The two different types of inactive
domains are shown in black and grey.
The active sites between the domains
are represented by white pixels. }
\label{Simulation}
\end{figure}}
%
%
%
%
%
\def\figresultsone{
\begin{figure}[tbp]
\begin{center}~
\epsfxsize=90mm
\epsffile{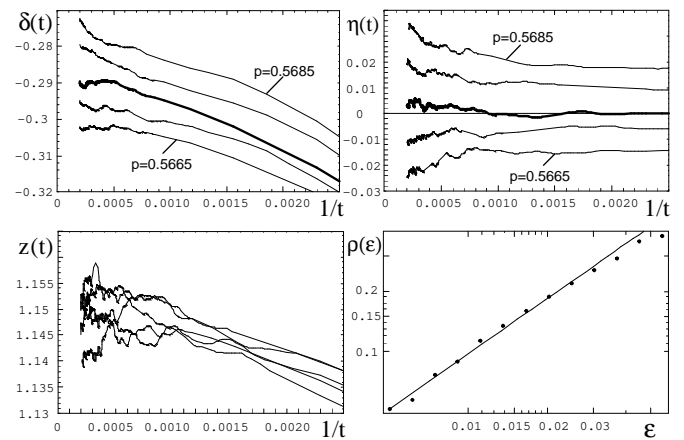}
\end{center}\vspace{-1mm}
\caption{Numerical results for model I. The effective exponents
$\delta(t)$, $\eta(t)$ and $z(t)$ are obtained from dynamic simulations
for $p=q=0.5665,0.5670,\ldots,0.5685$. The densitiy of active sites
$\rho(\epsilon)$ is measured in static simulations for different values
of $\epsilon=p-p_c$. The slope of line in the log-log plot gives an estimate
for the exponent $\beta$.}
\label{ResultsModelI}
\end{figure}}
%
%
%
%
%
\def\figresultstwo{
\begin{figure}[h]
\begin{center}~
\epsfxsize=90mm
\epsffile{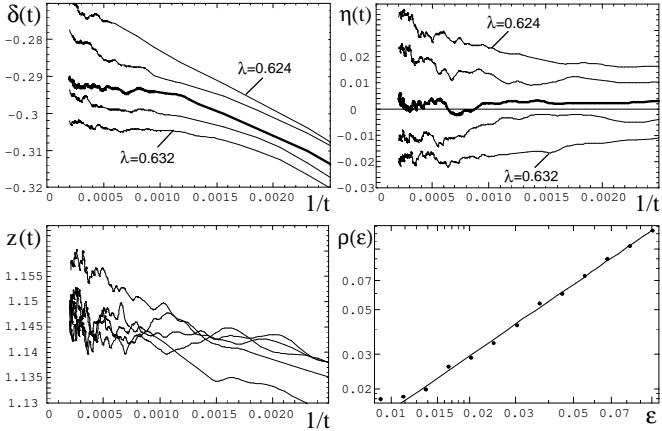}
\end{center}\vspace{-1mm}
\caption{Results of analog simulations of model II. The parameters
vary in the range $\lambda=\mu=0.624,0.626,\ldots,0.632$.}
\label{ResultsModelII}
\end{figure}}
%
%
%
%
%
\def\figsymmbreak{
\begin{figure}[h]
\begin{center}~
\epsfxsize=90mm
\epsffile{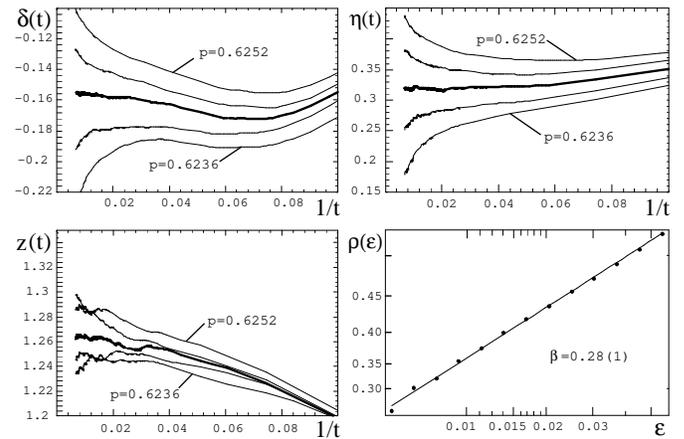}
\end{center}\vspace{-1mm}
\caption{Symmetry breaking field: Simulation of model I for slightly different
growth rates ($p_{1,2}=q\pm0.02$). DP exponents are recovered.}
\label{FigResultsSymmbreak}
\end{figure}}
%
%
%
%
%
\def\figtwodim{
\begin{figure}[h]
\begin{center}~
\epsfxsize=70mm
\epsffile{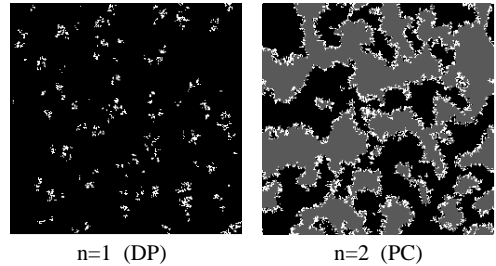}
\end{center}\vspace{-1mm}
\caption{Simulation snapshot of two-dimensional systems with
one and two absorbing states in the active phase near criticality. 
White dots denote active sites. }
\label{TwoDimensionSimul}
\end{figure}}
%
%
%
\def\Nexta   {   \hline        && \\[-4mm]}
\def\Next    {\\ \hline        && \\[-4mm]}
\def\Nextb   {\\ \hline \hline && \\[-4mm]}
\def\Nextc   {\\ \hline \hline && \\}
\def\NextEnd {\\ \hline}
\def\tableresults{

\begin{center}
\begin{tabular}{||c||c|c|c|c||}
\hline
\multicolumn{5}{||c||}{Known models:} \\ \hline
& $\delta$ & $\eta$ & $z$ & $\beta$ \\ \hline\hline

A and B \cite{Grassberger84} &
$0.27(8)$ & $-$ & $-$ & $0.6(2)$ \\ \hline

BAW $n=2$ \cite{Grassberger89} &
$0.283(16)$ & $0.272(12)^*$ & $1.11(2)$ & $0.94(6)$ \\ \hline

BAW $n=2$ \cite{Zhong} &
$0.285(2)$ & $0.000(1)$ & $1.141(2)$ & $0.92(3)$ \\ \hline

BAW $n=4$  \cite{Jensen94} &
$0.286(2)$ & $0.000(1)$ & $1.147(4)$ & $0.922(5)$ \\ \hline

kinetic Ising \cite{Menyhard} &
$0.27(2)$ & $0.30(2)^*$ & $1.14(2)$ & $0.80(8)$ \\ \hline

dyn. BAW  \cite{Kwon} &
$0.287(1)$ & $0.000(3)$ & $1.155(5)$ & $-$ \\ \hline

MDM \cite{MonomerDimer} &
$0.29(2)$ & $0.00(2)$ & $1.34(20)$ & $0.88(3)$ \\
\hline
\multicolumn{5}{||c||}{Present work:} \\ \hline

model I &   $0.285(10)$ & $0.00(1)$ & $1.15(1)$ & $0.90(5)$  \\ \hline

model II &   $0.29(1)$ & $0.00(1)$ & $1.15(1)$ & $0.93(5)$ \\ \hline
\end{tabular}
\end{center}
{\small
\begin{center}
TABLE I. Critical exponents of models in the PC class. The star indicates
values measured in kink dynamics where a different scaling relation holds.
\end{center}}
}
%
%
%
\def\figsandcaptions{}
%
%
%
%
\section{Introduction}
The study of stochastic lattice models exhibiting a
continuous phase transition from an active phase into an
absorbing state is a field of growing interest. 
In these models the dynamical processes 
take place close to an absorbing state, 
i.e. a configuration once reached, the system cannot escape from.
Most of them belong to the universality class of
directed percolation (DP); the best known examples are
DP lattice models \cite{DirectedPercolation},
the Domany-Kinzel cellular automaton \cite{DomanyKinzel}, the
contact process \cite{ContactProcess},
Schl\"ogl's first and second model \cite{Schloegl},
and branching annihilating walks with an odd
number of offspring \cite{BAWOdd}. 
In a field-theoretical formulation these models
can be related to Reggeon field theory~\cite{RFT},
which was proven to be in the same
universality class as directed percolation \cite{RFTProof}.

The variety of DP models led Janssen and Grassberger 
to the conjecture that in one-component models 
all continuous phase transitions from
an active phase to a {\it single} absorbing 
state  are in the DP universality 
class~\cite{SingleAbsorbingState}. 
However, the known examples for DP include 
even more complicated systems, e.g.,
multi component systems \cite{DPMultiComponent} and models with
several absorbing states \cite{DPMultiAbsState}. 
Some models that were initially thought to be in different
classes were later found to belong to the DP class as well
\cite{AlsoInDP}. Thus the directed percolation universality class
is extremely robust and covers a wide range of models. 

Among the models with absorbing states only a few exceptions are
known that do not belong to the DP universality class. During the 
past few years it became clear that they represent a universality class 
that is different from that of directed percolation. The known examples
are the models A and B of probabilistic cellular automata
\cite{Grassberger84,Grassberger89}, nonequilibrium kinetic
Ising models with combined zero- and infinite-temperature dynamics
\cite{Menyhard}, interacting monomer-dimer models \cite{MonomerDimer},
and branching annihilating walks (BAW's) with an even number of offspring
\cite{BAW2PhaseTrans,Jensen93,Jensen94,Zhong,CardyFieldTheory}. 
The common feature of all these models is that 
the number of particles (or kinks) is conserved mod 2.
Therefore the new class is sometimes referred to as the
{\it parity-conserving} (PC) class.

Initially parity conservation was thought to be the reason for the emergence
of the different universality class. 
However, Park and Park \cite{ParkSymmBreak}
recently showed that in the example of an interacting monomer-dimer model
a weak parity-conserving external field can force the system back to
the DP class. They concluded that the essential property of the class is
not parity conservation, but a symmetry among different absorbing states.

In order to address this question,
we propose two one-dimensional models that exhibit
a phase transition from an active to an inactive phase
consisting of $n$ equivalent absorbing states. These models
generalize two well-known stochastic models for directed percolation, 
namely, the Domany-Kinzel cellular automaton \cite{DomanyKinzel}
and the contact process \cite{ContactProcess}. We can conclude from numerical
simulations that for $n=2$ both models belong to the PC class.

The models we define
are interesting for various reasons. As generalizations of
well-known DP models they give a better physical understanding of models in 
the PC class. Moreover, they are defined by ordinary two-site 
nearest-neighbor interactions (rather than three- or four-site interactions) 
and can be generalized easily to both higher dimensions and a higher 
number of absorbing states. In addition, unlike previously known models, 
our models do not explicitly conserve parity. 
This confirms that the symmetry among the absorbing states is indeed the only
essential property of models in the PC class.

Before defining the models let us present an intuitive idea how DP models
can be generalized to models with several absorbing states. 
Directed percolation models are usually defined
on some $d$-dimensional lattice whose sites can be either active (wet)
or inactive (dry). If all sites are inactive, the system is in an
absorbing configuration from that it cannot escape.
In the presence of active sites, the system evolves 
in time according to specific local processes. Although microscopically
these processes  may be defined differently, 
most of the DP models have the feature in  common that 
their time evolution seen on a large scale is subject to the following rules:
$$ \begin{array}{ll}
\mbox{\bf (a)} &
\mbox{Inactive (dry) spots are created randomly
within} \\ &
\mbox{active (wet) islands.} \nonumber\\[3mm]
\mbox{\bf (b)} &
\mbox{The boundaries between active and inactive} \\ &
\mbox{domains fluctuate in a way that active islands}
\nonumber \\ &
\mbox{are biased to grow.}
\end{array} $$
Both processes {\bf (a)} and {\bf (b)} compete with each other. Depending on
their probabilities the system can be in two different phases.
If the probability for {\bf (a)} is very small,
the system is in the {\it active phase} where,
starting with a nonzero density of active sites, active clusters
percolate constantly. If the probability for
{\bf (a)} is very large, the system is in the {\it inactive phase} where
active clusters die out wherefore eventually the system enters
the absorbing state. At the percolation threshold the system 
goes through a continuous phase transition where
the details of the local processes become irrelevant 
and long-range fluctuations can be observed. 
According to the DP conjecture \cite{SingleAbsorbingState}, we assume
that all models with a single absorbing state
defined in the spirit of rules {\bf (a)} and {\bf (b)} 
belong to the universality class of directed percolation.

The main idea of the present work is that a generalization of the
above rules to $n$ equivalent absorbing states generates 
universality classes different from DP, 
in particular to the PC class in the case of two symmetric absorbing states.
Such a generalization can be defined as follows.
Let us assume that each inactive site carries a 
``color'' labeled by $1,\ldots,n$.
The simplest generalization of the rules {\bf (a)} and {\bf (b)} is
the following:
$$ \begin{array}{ll}
\mbox{\bf (A)} &
\mbox{Inactive spots of random color $1,\ldots,n$ are created} \\ &
\mbox{randomly within active islands. } \nonumber\\[3mm]
\mbox{\bf (B)} &
\mbox{The boundaries between active and inactive} \\ &
\mbox{domains fluctuate in a way that active islands}
\nonumber \\ &
\mbox{grow.}\\[3mm]
\mbox{\bf (C)} &
\mbox{Boundaries between inactive domains of different} \\ &
\mbox{colors are not allowed to stick to each other}
\nonumber \\ &
\mbox{irreversibly. They are free to separate again} \\ &
\mbox{leaving active sites in between.}\end{array} $$
Rule {\bf (A)} and {\bf (B)} are straightforward generalizations
of {\bf (a)} and {\bf (b)}. Again both processes 
compete with each other and lead to a phase 
transition from an active to an inactive phase. 
Rule {\bf (C)} is different and distinguishes the
different colors. Roughly speaking, this rule tells us that
between two inactive domains of different colors a thin film of
wet sites is preserved. The importance of this rule becomes 
obvious by looking at the contrary: If domains of different 
colors were allowed to stick to each other irreversibly, the colors would
then be irrelevant. This would mean that the process
is compatible with the previous rules {\bf (a)} and {\bf (b)}
and thus belong to directed percolation.
Rule {\bf (C)} allows wet sites between
absorbing domains of different colors 
to survive for a long time. This slows down the relaxation 
towards one of the absorbing states and therefore we expect 
systems with several absorbings states to be ``more active''
than usual DP models.

Another important requirement is that 
the rules are {\it symmetric}
under global permutation of the colors. 
We will show that if this symmetry is broken,
one of the colors begins to play a dominant role so that the phase
transition is again in the DP universality class.
Although these rules give only an intuitive description rather that a strict
definition, they will help us to define models with several
absorbing states.


\section{Definition of the models}
\subsection{Model I: Generalized Domany-Kinzel cellular automaton}
In the Domany-Kinzel model \cite{DomanyKinzel} the state at 
a given time $t$ is specified by binary variables
$\{s_i\}$, which can have the values $A$ (active) and $I$ (inactive).
At odd (even) times, odd- (even-) indexed sites are updated according
to specific conditional probabilities. This defines a cellular automaton
with parallel updates (discrete time evolution) acting on two
independent triangular sublattices:
\figdomanykinzel
The conditional probabilities 
in the Domany-Kinzel model
$P(s_{i,t+1} \,|\, s_{i-1,t}\,,\,s_{i+1,t})$
are given by
\begin{eqnarray}
\label{eq1}
P(I\,|\,I,I)&=&1\,,\\
\label{eq2}
P(A\,|\,A,A)&=&q\,,\\
\label{eq3}
P(A\,|\,I,A)=P(A|A,I)&=&p\,,
\end{eqnarray}
and $P(I|s_{i-1},s_{i+1})+P(A|s_{i-1},s_{i+1})=1$,
where $0\leq p\leq 1$ and $0\leq q\leq 1$
are two parameters. Equation (\ref{eq1}) ensures that the configuration
$\ldots,I,I,I,\ldots$ is the absorbing state.
The process in Eq. (\ref{eq2}) corresponds to rule~{\bf (a)} 
and describes the creation of inactive (dry) spots within 
active (wet) islands with probability $1-q$.
The  random walk of boundaries between active and inactive domains
is realized by the processes in Eq.~(\ref{eq3}).
According to rule {\bf (b)},
DP transitions can be observed only if $p>\frac12$
when active (wet) islands are biased to grow.
The processes and their probabilities can be summarized
in the form of a probability table:
\begin{center}
\begin{tabular}{||c||c|c||}
\hline
$s_{i-1},s_{i+1}$ & $P(A\,|\,s_{i-1},s_{i+1})$ & 
$P(I\,|\,s_{i-1},s_{i+1})$  \\ \hline\hline
$AA$ & $q$ & $1-q$ \\ \hline
$AI$ & $p$ & $1-p$ \\ \hline
$IA$ & $p$ & $1-p$ \\ \hline
$II$ & $0$ & $\,\,\,1\,\,\,$\\ \hline
\end{tabular}
\end{center}
We now define a generalization of the Domany-Kinzel model
following the rules {\bf (A)}--{\bf (C)} (hereafter referred to as model I).
This model has $n+1$ states per site: one active state $A$ 
and $n$ different inactive states $I_1,I_2,\ldots,I_n$.
The conditional probabilities are given by
$(k,l=1,\ldots,n; \,\, k\neq l)$
\begin{eqnarray}
\label{eqa}
P(I_k\,|\,I_k,I_k)&=&1\,,\\
\label{eqb}
P(A|A,A)=1-n\,P(I_k|A,A)&=&q\,, \\
\label{eqc}
P(A|I_k,A)=P(A|A,I_k)&=&p_k\,,  \\ \nonumber
P(I_k|I_k,A)=P(I_k|A,I_k)&=&1-p_k\,,\\
\label{eqd}
P(A|I_k,I_l)&=&1\,, 
\end{eqnarray}
where we study the symmetric case $p_1,\ldots,p_n=p$.
Equations (\ref{eqa})--(\ref{eqc}) are straightforward
generalizations of Eqs. (\ref{eq1})--(\ref{eq3}).
The only different process 
is the creation of active sites between
two inactive domains of different colors
in Eq.~(\ref{eqd}) according to  rule {\bf (C)}. 
For simplicity we chose the 
probability of this process to be equal to one. 
We may also use a probability less than one, but
it turned out that this does not change
the critical properties of the system.
\\
\indent
For $n=1$ the model defined above reduces to the original Domany-Kinzel
model. In Sec. \ref{SimulSection} we will investigate the
generalized Domany-Kinzel model with $n=2$ absorbing states. 
The corresponding probability table reads:
\begin{center}
\begin{tabular}{||c||c|c|c||}
\hline
$s_{1},s_{2}$ & $P(A  \,|\,s_{1},s_{2})$ & 
$P(I_1\,|\,s_{1},s_{2})$ & $P(I_2\,|\,s_{1},s_{2})$ \\ \hline\hline
$AA$     & $q$ & $1-q/2$ & $1-q/2$ \\ \hline
$AI_1$   & $p$ & $1-p$   & $0  $   \\ \hline
$AI_2$   & $p$ & $0$     & $1-p$   \\ \hline 
$I_1A$   & $p$ & $1-p$   & $0$     \\ \hline 
$I_2A$   & $p$ & $0$     & $1-p$   \\ \hline
$I_1I_1$ & $0$ & $1$     & $0$     \\ \hline 
$I_1I_2$ & $1$ & $0$     & $0$     \\ \hline 
$I_2I_1$ & $1$ & $0$     & $0$     \\ \hline 
$I_2I_2$ & $0$ & $0$     & $1$     \\ \hline
\end{tabular}
\end{center}
\subsection{Model II: Generalized contact process}
The one-dimensional contact process is the simplest example for
a DP model with continuous time evolution~\cite{ContactProcess}.
Its dynamics is defined by nearest-neighbor processes that occur
spontaneously due to specific rates (rather than probabilities).
In numerical simulations models of this type are usually realized
by random sequential updates. This means that
a pair of sites $\{s_i,s_{i+1}\}$ is chosen at random and an update is
attempted according to specific transition rates
$w(s_{i,t+dt}\,,\,s_{i+1,t+dt}\,|\,
s_{i,t}\,,\,s_{i+1,t})$. 
\figcontactprocess
Each attempt to update a pair of sites increases
the time $t$ by $dt=1/N$, where $N$ is the total number of sites. One
time step (sweep) therefore consists of $N$ such attempts. The
contact process is defined by the rates
\begin{eqnarray}
\label{gl1}
w(A,I\,|\,A,A)=w(I,A\,|\,A,A)&=&\lambda\,,\\
\label{gl2}
w(I,I\,|\,A,I)=w(I,I\,|\,I,A)&=&\mu\,,\\
\label{gl3}
w(A,A\,|\,A,I)=w(A,A\,|\,I,A)&=&1\,,
\end{eqnarray}
where $\lambda>0$ and $\mu>0$ are two parameters (all other rates are zero).
Equation (\ref{gl1}) describes the creation of inactive (dry) spots
within active (wet) islands corresponding to rule {\bf (a)}.
Equations (\ref{gl2}) and (\ref{gl3}) describe the shrinkage and 
growth of active islands according to rule {\bf (b)}. In order to fix 
the time scale, we chose the rate in Eq. (\ref{gl3}) to be equal to one. 
The active phase is restricted to the region $\mu < 1$ where
wet islands are likely to grow.

As in the case of the Domany-Kinzel model,
we define a generalization of the contact process by introducing
$n$ different inactive states $I_1, I_2,\ldots,I_n$. The dynamics of
the generalized model (model II) is defined by the rates 
\begin{eqnarray}
\label{ans1}
w(A,I_k\,|\,A,A)=w(I_k,A\,|\,A,A)&=&\lambda/n\,,\\
\label{ans2}
w(I_k,I_k\,|\,A,I_k)=w(I_k,I_k\,|\,I_k,A)&=&\mu_k\,,\\
\label{ans3}
w(A,A\,|\,A,I_k)=w(A,A\,|\,I_k,A)&=&1\,,\\
\label{ans4}
w(I_k,A\,|\,I_k,I_l)=w(A,I_l\,|\,I_k,I_l)&=&1\,,
\end{eqnarray}
where $k,l=1,\ldots,n$ and $k\neq l$ (all other
rates are zero). Again we consider the
symmetric case $\mu_1,\ldots,\mu_n=\mu$.
Eqs. (\ref{ans1})--(\ref{ans3}) are generalizations
of Eqs. (\ref{gl1})--(\ref{gl3}). The new rule {\bf(C)} is implemented by
Eq. (\ref{ans4}) which describes the creation of active sites between 
two inactive domains of different colors. 
For $n=1$ the model defined above is reduced to the usual contact process
(\ref{gl1})--(\ref{gl3}).

\figphasediag

\subsection{Phase diagrams}
The phase diagrams of both models are shown in Fig.~\ref{PhaseDiagram}.
The active (inactive) phase is characterized by a nonzero (vanishing)
density of active sites in the thermodynamic limit.
Both phases are separated by a phase transition line (the bold line in
Fig. \ref{PhaseDiagram}). The dashed line indicates the corresponding 
phase transition for a single absorbing state. Comparing both lines we notice 
that generally models with two absorbing states tend to be 'more active' 
than their DP counterparts (for exceptions see Ref. \cite{Kwon}).

We checked numerically that as in DP the critical exponents of 
the generalized models are the same all along the phase transition line. 
The only exceptions are the ending points $(p,q)=(1/2,1)$ and 
$(\lambda,\mu)=(0,1)$, where the transition lines for $n=1$ and $n=2$ 
intersect (marked by A in Fig. \ref{PhaseDiagram}).
In these points rule {\bf (A)} is no longer valid and the 
entirely active configuration $\ldots AAA \ldots$ emerges as an additional 
absorbing state. This leads to a different universality class, which, in
the case of $n=1$, is referred to as 'compact directed percolation'. 


\subsection{General properties}

A typical time evolution of models with two absorbing states is 
shown in Fig. \ref{Simulation}. In the active phase
$(\lambda<\lambda_c)$ small inactive islands of 
random color are generated and exist only briefly. 
Approaching the phase transition their size and 
lifetime grows while the density of active sites decreases.
Notice that according to rule {\bf (C)} a thin film of active sites 
separates different inactive domains.

An important property of models with several absorbing 
states is a very different relaxation
towards the absorbing configuration. For DP in the inactive 
phase the order parameter $\rho$ is known to decay 
{\it exponentially} in time. However, this is not true for 
models with two absorbing states.  As shown in Fig. \ref{Simulation} 
$(\lambda>\lambda_c)$, starting from a random initial configuration, large 
domains of different colors are formed. These domain walls survive and 
diffuse until they annihilate mutually. In this annihilation process 
the density of domain walls is known to decay 
{\it algebraically} like $\rho(x) \sim t^{-1/2}$
\cite{Annihilation}. Because of the slow relaxation 
numerical simulations of models with several absorbing 
states are more difficult to perform.

\figsimulations

It should be emphasized that in the models defined above there
is no explicit parity conservation on the microscopic level:
In each local update no more than one site is modified [cf.
Eqs. (\ref{ans1})--(\ref{ans4})]. Therefore it is impossible to create
more than two kinks or particles per update (a nontrivial paritiy-conserving 
dynamics requires the generation of at least three kinks
$X\rightarrow 3X$, $2X\rightarrow 0$). Nevertheless, the annihilating
domain walls described above, by their very nature, obey a 
parity-conserving dynamics. Therefore parity conservation can still be 
seen on large scales. We will return to this observation in Sec. 
\ref{BAWDiscussion}.


\section{Two symmetric absorbing states: Numerical results}
\label{SimulSection}
\subsection{Monte Carlo simulations}
%
In order to measure the critical exponents of models I and II in the case
of two absorbing states, we perform dynamic Monte Carlo simulations
(see, e.g., \cite{Zhong}). We use defect dynamics, i.e., we start with an 
initial configuration where all sites are in the inactive state $I_1$
except for one active site in the center. The system then evolves along the
dynamic rules of the model. 
For various values of the parameters near point B in Fig. \ref{PhaseDiagram}
we perform $10^6$ independent runs up to $5000$ time steps.
However, most of them stop earlier because the system enters into the
absorbing configuration where all sites are in the state 
$I_1$\footnote{In an infinite system, there is no way to
reach the other absorbing configuration $I_2$.}. 
In order to avoid finite-size effects, we adjust the
system size after each time step 
according to the actual size of the active cluster.
As usual in this type of simulations,
we measure the survival probability $P(t)$, the number of active sites $N(t)$,
and the mean square of spreading  from the
origin $R^2(t)$ averaged over active runs. 
At criticality, these quantities are expected to scale 
algebraically in the long-time limit
\begin{equation}
\label{quantities}
P(t) \sim t^{-\delta} \,, \hspace{10mm}
N(t) \sim t^\eta \,, \hspace{10mm}
R^2(t) \sim t^z\,.
\end{equation}
\figresultsone 
\figresultstwo
The critical exponents are related to the exponents $\beta$,
$\nu_\perp$, and $\nu_{||}$ by
\begin{equation}
\delta = \frac{\beta}{\nu_{||}}\,,
\hspace{10mm}
z = \frac{2 \nu_\perp }{\nu_{||}}
\end{equation}
and obey the scaling relation
\begin{equation}
4\delta \,+\, 2\eta \;=\; d\,z\,.
\end{equation}
The quantities (\ref{quantities})
show straight lines in double logarithmic plots. Off criticality,
the lines are curved. In order to get precise estimates for the scaling
exponents, it has been useful to consider the local slopes of the curves by
introducing the {\it effective exponents}
\begin{equation}
-\delta(t) \;=\; \frac{\log_{10}\bigl(P(t)/P(t/b)\bigr)}{\log_{10} b}
\end{equation}
and similarly $\eta(t)$ and $z(t)$, where $\log_{10} b$ 
is the distance used for
estimating the slope. Choosing $b=5$, we measured the effective
exponents of both models for various values of $q=p$ and $\lambda=\nu$.
The results of our simulations are shown in Figs.
\ref{ResultsModelI} and \ref{ResultsModelII}.
Off criticality, the curves for $\delta(t)$ and $\eta(t)$ show negative or
positive curvature. The figures give us an estimate of
the critical points $p_c=0.5673(5)$
for model I and  $\lambda_c=0.628(2)$ for model II.
The estimates for the critical exponents are
$\delta=0.285(10)$, $\eta=0.00(1)$, and $z=1.15(1)$ for model I and
$\delta=0.29(1)$, $\eta=0.00(1)$, and $z=1.15(1)$ for model~II.

The exponent $\beta$ has been obtained directly in 
static simulations by measuring the steady state density $\rho$
in the active phase near the critical point. Although this method is
known to be quite inaccurate to determine $\beta$ and 
we measured only over one decade in $\epsilon$, we
found the reasonable values $\beta=0.90(5)$ for model~I and
$\beta=0.93(5)$ for model~II.

The estimates of the critical exponents agree with previous
results for models in the PC class (cf. Table I).
Thus, from our numerical results we can conclude that
for $n=2$ both models belong to the PC universality class.
\tableresults
\subsection{Symmetry breaking field}
Recently Park and Park showed in the example of an interacting 
monomer-dimer model that if the symmetry among the absorbing statess
is broken by an external field, 
the DP universality class is recovered~\cite{ParkSymmBreak}.
In order to verify this observation, we introduce an external field 
by modifying the growth rates for inactive 
islands of different colors. This can be done by choosing different $p_k$ 
in Eq. (\ref{eqc}) [$\mu_k$ in Eq. (\ref{ans2})].
Because of the different growth rates, one of the colors is going
to play a dominant role wherefore in the large-scale limit the 
system evolves as if it had only a single absorbing state. 

\figsymmbreak

To demonstrate this, we repeat the above simulations of model I 
for $p_{1,2}=q \pm 0.02$. The critical point is shifted to $q_c=0.6245(10)$. 
The exponents we measure (cf. Fig. \ref{FigResultsSymmbreak})
are $\delta=0.16(2)$, $\eta=0.32(4)$, $z=1.27(4)$,
and $\beta=0.28(1)$, which agree with the DP exponents  \cite{Jensen94}
$\delta=0.159(1)$, $\eta=0.314(1)$, $z=1.266(2)$, and $\beta=0.2765(1)$.

If the symmetry of a model with $n>2$ absorbing states is partially broken,
a subset of $m$ colors starts to play a dominant role. We expect that such
a system behaves at criticality like a model with $m$ absorbing states.
%
%
%
\section{Other special cases}

\subsection{Compact clusters}
As mentioned before, the Domany-Kinzel model and the contact process have
a line in their phase diagram where the entirely active configuration
emerges as an additional absorbing state. On this line active clusters 
are not fractal but compact, wherefore it is called the compact directed 
percolation line. Here the dynamical processes are exactly solvable 
by reducing it to an annihilation-diffusion process of kinks $2X \rightarrow 0$. 
Although this leads to a  different universality class at the phase transition 
point  (point A in Fig. \ref{PhaseDiagram}), compact DP has been used in 
many cases to improve the understanding of ordinary DP.

A similar situation exists in generalized models with
several absorbing states. The fact that the transition A point is 
identical to that of ordinary compact DP indicates that this point is
exactly solvable in all cases. However, the translation into a 
kink language is slightly more complicated. 
Since kinks between inactive domains of different colors cannot exist
[rule {\bf (C)}], only $n$ types of kinks $X_k$ between active and inactive 
domains ($AI_k$, $I_kA$) play a role. 
These kinks have to occur in pairs and undergo an 
annihilation-diffusion process $2X_k \rightarrow 0$ 
(kinks of different types cannot
annihilate). It is important that there is no more generation of
randomly colored inactive domains. Therefore dynamical simulations such as 
those in the previous section starting from an initial configuration with
only one type of inactive sites yield the same results for all $n$, 
i.e., $\delta=1/2$ and $z=1$. However,  compact percolation processes 
are known to depend strongly on initial conditions so that generally the 
situation may be more complicated.

\subsection{Simulations in higher dimensions}
\label{SectHigherDim}
%
The models presented in this paper
can easily be generalized to higher dimensions.
This is particularly simple in the case of model II since its definition
(\ref{ans1})--(\ref{ans4}) 
can be used in any dimension. In simulations we observed that for $d=2$
this system has a phase transition although the relaxation towards
one of the absorbing states in the inactive phase is extremely slow. 
Figure \ref{TwoDimensionSimul} shows typical configurations in the active 
phase near criticality. 
While in ordinary DP active clusters are separated spatially, 
active sites in two-dimensional models with two absorbing states
are arranged in fractal ``lines'' along the boundaries of 
inactive islands. Repeating the simulations described above 
on a $80 \times 80$ lattice we obtained 
$\lambda_c=0.99(1)$, $\delta=0.9(1)$, $\eta=0.00(5)$, and $z=1.0(1)$. 
These results agree roughly with the mean-field exponents
$\delta=1, \eta=0, z=1$ and $\beta=1$ (see next Section). 
Therefore, we conjecture that $1 < d_c \leq 2$ is the upper critical
dimension of systems with two absorbing states.
\figtwodim

\subsection{Relation to BAW's}
%
%
\label{BAWDiscussion}
We already mentioned that models I and II have the same critical
behavior all along the phase transition line (except point A
in Fig. \ref{PhaseDiagram}). Moving along this line we can control the
mean size of active islands, which is infinite at point A, 
a few sites at point B, and one site at point C (in model II
the latter case corresponds to taking $\lambda \rightarrow \infty$).
In one dimension point C can be related to branching walk models
since each active site can be interpreted as a single walker. For
$n=1$ these walkers diffuse and interact by $2X \leftrightarrow X$,
$X \rightarrow 0$, which results in a DP process.
For $n \geq 2$ the situation is more complicated. Let us denote by $X_{jk}$
a walker separating inactive domains $I_j$ and $I_k$. Then the processes
are $X_{jk} \leftrightarrow X_{jl}X_{lk}$ (with random $l$)
and  $X_{jj} \rightarrow 0$. These processes clearly do not conserve parity.
However, because of $X_{jj} \rightarrow 0$ only walkers between
domains of different color $j \neq k$ survive for a long time
[cf. rule {\bf (C)}]. Hence for $n=2$ the majority of walkers 
react like
\begin{eqnarray}
&&X_{12} \rightarrow X_{11}X_{12} \rightarrow X_{12} \nonumber \\
&&X_{12} \rightarrow X_{11}X_{12} \rightarrow X_{12} X_{21} X_{12} \nonumber \\
&&X_{12}X_{21} \rightarrow 0 \nonumber \\
&&\mbox{with analog reactions for $X_{21}$} \nonumber
\end{eqnarray}
Thus, in a long time limit the walkers undergo an effective reaction of the type
$X \rightarrow 3X$, $2X \rightarrow 0$, which is a BAW with two offspring 
in one dimension.

The relation to BAW modes is even more general and holds not only at
point C but everywhere on the phase transition line except for point A.
The walkers then have to be identified with domain walls seperating 
inactive domains of different colors. At first glance this
seems to be contradicting: Domain walls, by their very definition, obey
a local parity-conserving dynamics. On the other hand, it is obvious that
the local processes do not conserve parity. However, domain walls 
in our models are extended objects. Their thickness fluctuates and
varies along the phase transition line from typically one site 
at point C up to infinitely many sites at point A. It is important
to notice that the domain walls simply cannot be identified with active
islands (which may also occur between inactive domains of the same color)
but require a more complicated definition. Although there is no microscopic
parity conservation, a careful analysis shows that the dynamical rules 
ensure that all microscopic processes violating parity conservation 
have a very short lifetime. This is the reason why an effective 
parity -conserved dynamics can be recovered in the limit 
of large scales in time and space. 

In higher dimensions $d \geq 2$ the physical properties of BAW's with
an even number of offspring are governed by the mutual annihilation of
the walkers \cite{ReviewDickman}. However, the models presented in this 
paper behave very differently in higher dimensions, which 
makes it impossible to relate them to BAW's.
As shown in Fig. \ref{TwoDimensionSimul}, the active sites in models
I and II arrange themselves as $(d-1)$-dimensional surfaces separating
inactive domains of different colors. Thus they cannot be interpreted
as pointlike random walkers. In $d \geq 2$ dimensions we therefore 
expect BAW's with an even number of offspring to be in a different
universality class from the present models.
\subsection{More than two absorbing states}
No results were obtained for $n\geq 3$ symmetric absorbing states
in one dimension. It turned out that it is 
impossible to determine the critical point because the plots for 
$\delta(t)$, $\eta(t)$, and $z(t)$ show only one type of curvature. 
This observation agrees with recent results obtained for an $N$-species
generalization of BAW models with an even number of offspring
\cite{CardyFieldTheory}. For $N \geq 2$ these models are always 
in the active phase and their critical exponents are described by yet another 
universality class.
\subsection{Mean-field approximation}
Denoting by $\rho_k$ the density of inactive sites $I_k$, the processes
in model II imply the mean-field equation
\begin{equation}
\frac {\partial}{\partial t} \rho_k \;=\;
\frac{2 \lambda}{n} \rho_0^2 + 
\mu \rho_0 \rho_k +
\rho_k^2 - \rho_k\,,
\end{equation}
where $\rho_0 = 1-\sum_{j=1}^n \rho_j$. Choosing $\lambda=\mu$, the
critical point is $\lambda_c=1$ for $n=1$ and $\lambda_c=\infty$
for $n \geq 2$, which means that in mean-field models with more than 
two absorbing states are always in the active phase 
(see also Ref. \cite{CardyFieldTheory}).
For all $n$ the density of active sites $\rho_0$ scales
like  $(\lambda-\lambda_c)^\beta$ with the mean-field exponent $\beta=1$.
This is close to the measured values $\beta \sim 0.9$ in the model with
two absorbing states, which indicates that we are already close to the
upper critical dimension.

\section{Summary and discussion}
\label{Summary}
We have shown by the example of the Domany-Kinzel model and the contact
process that lattice models for directed percolation can be generalized
to models with $n$ symmetric absorbing states. Numerical simulations
lead to the conclusion that such models in one dimension with two
absorbing states belong to the PC universality class. Since these models
do not explicitly conserve any quantity mod 2 they show that rather than
parity conservation the symmetry among the absorbing states is the 
origin for the emergence of a different 
class. As soon as this symmetry is broken,
the critical behavior jumps back to DP.

The symmetry used in our models is the group of permutations $S_n$.
No reliable numerical results could be obtained for $S_3$ and
higher symmetries. However, one may also introduce other symmetries 
such as cyclic groups (e.g., $Z_3$) and investigate whether they define 
different universality classes 
(cyclic symmetries of this type appear, e.g., in the
three-candidate voter model \cite{Tainaka}). It would be also 
interesting to examine models with more than one symmetric active states.

For a better understanding it would be desirable to find an appropriate
field-theoretical description of the model. For BAW's with an even
number of offspring this has been done recently in Ref.
\cite{CardyFieldTheory}. There are surprising results, in particular
one has two different critical dimensions, one of them ($d_{c_1}=4/3$) 
related to the properties of active clusters and the other ($d_{c_2}=2$) 
related to the annihilation process. Although the application of this
theory to the present type of model may not be transparent, Cardy
and T\"auber were able to identify an $S_2$ symmetry on an operator level.
This again indicates that this symmetry plays an important role 
in the PC class.

A field theory for the present type of model should be different from that of
BAW's because of its different phenomenology in higher dimensions (see
Secs. \ref{SectHigherDim} and \ref{BAWDiscussion}). 
Nevertheless, both theories 
should give identical results in  one dimension. 
The simplest ansatz for such a field theory would be to add a
diffusion like term and $n$ noise fields to the mean-field equations
\begin{eqnarray}
\frac {\partial}{\partial t} \rho_k(x,t) &=&
\frac{2 \lambda}{n} \rho_0^2(x,t) + 
\mu \rho_0(x,t) \rho_k(x,t) +
\rho_k^2(x,t) \nonumber \\
&&- \rho_k(x,t) + D\,\nabla^2 \rho_k(x,t) + \eta_k(x,t)\,.
\end{eqnarray}
However, even if this were correct, the derivation of the correlations
in the noise remains a highly nontrivial problem. Therefore, the development
of appropriate field theories is a challenge towards
a better understanding of universality classes appearing in 
systems with several absorbing states.
%
%
%
%
\\[10mm]
{\bf Acknowledgments}\\[1mm]
I would like to thank U. Alon, E. Domany, M. Evans, Y. Goldschmidt,
A. Honecker, D. Mukamel, Z. R\'acz, and S. Sandow for helpful discussions.
This work was supported by the Minerva Foundation.
%
%
%
%
%
%

%
%
\figsandcaptions
\end{document}